\documentclass[a4paper,floatfix,superscriptaddress,aps,prl,twocolumn,showpacs,notitlepage,nofootinbib,nobibnotes]{revtex4-2}
\usepackage[utf8]{inputenc}
\usepackage[T1]{fontenc}
\usepackage[sc,osf]{mathpazo}
\usepackage{amsmath, amsthm, amssymb, amsfonts, mathbbol, amstext}
\usepackage{graphicx}
\usepackage{dcolumn}
\usepackage{bm}
\usepackage{bbm}
\usepackage{xcolor}
\usepackage[colorlinks=true,citecolor=blue,linkcolor=purple,urlcolor=blue]{hyperref} 
\usepackage{mathtools}

\usepackage{multirow}
\usepackage{siunitx}
\usepackage{physics}

\usepackage[normalem]{ulem}

\begin{document}
\title{A prescriptive method for fibre polarisation compensation in two bases}

\author{Teodor Strömberg}
\author{Peter Schiansky}
\author{Philip Walther}\affiliation{University of Vienna, Faculty of Physics \& Vienna Doctoral School in Physics, Boltzmanngasse 5, A-1090 Vienna, Austria}

\begin{abstract}
\textbf{Single-mode optical fibres exhibit a small but non-negligible birefringence that induces random polarisation rotations during light propagation. In classical interferometry these rotations give rise to polarisation-induced fading of the interferometric visibility, and in fibre-based polarimetric sensors as well as quantum optics experiments they scramble the information encoded in the polarisation state. Correcting these undesired rotations is consequently an important part of many experiments and applications employing optical fibres. In this Lab Note we review an efficient method for fully compensating fibre polarisation rotations for general input states. This method was not originally devised by us, but does to the best of our knowledge not appear in the literature, and our interactions with the community have indicated that it is not well known.}
\end{abstract}

\maketitle

\section{Introduction}
While single-mode (SM) optical fibres have a core that is nominally circularly symmetric, stress in the fibre can break this symmetry and give rise to a small amount of birefringence. The effect of this can be modelled as a series of birefrigent elements with random axes and retardances~\cite{imai1988polarization}, which even for fibre patch cables of moderate length can result in completely random transformations of an input polarisation state. Under the assumption that the total polarisation mode dispersion is negligible, these polarisation transformations can be considered unitary, and throughout the remainder of this text we will assume this to be the case.

Random polarisation rotations can have a detrimental effect on many applications of fibre optics, such as coherent receivers used in high-bandwidth telecommunication systems~\cite{smith1987techniques}, and interferometric fibre sensors~\cite{sheem1979polarization,stowe1982polarization}. This has led to the development of a variety of application-specific techniques for the control of the polarisation of light in fibres~\cite{okoshi1985polarization,ulrich1979polarization,kersey1991polarisation,kidoh1981polarization}, using both mechanical~\cite{lefevre1980single,shimizu1991highly} and electro-optical~\cite{shimizu1988endless,heismann1994analysis,chen2011broadband} fibre polarisation controllers. In the aforementioned contexts the problem of polarisation compensation consists of transforming a given, potentially time-varying, state of polarisation to a fixed state. This can be framed as a minimisation problem that can be satisfied using two quarter-wave plates to rotate the polarisation~\cite{reddy2016polarization}. In practice, however, devices additionally containing one half-wave plate, capable of realising any unitary polarisation rotation~\cite{SIMON1990165}, are more frequently employed. In some applications, such as fibre-optic gyroscopes~\cite{lefevre2022fiber}, polarisation maintaining (PM) fibres~\cite{noda1986polarization} are instead used to ensure a fixed output state.

In many quantum-information applications of fibre optics~\cite{flamini2018photonic}, such as quantum key distribution~\cite{gisin2002quantum}, as well as in certain classical and quantum sensors~\cite{kersey1988observation,yao2019polarimetry,silvestri2023experimental} the problem of polarisation compensation is more complicated. This is because quantum information encoded in a polarisation state is sensitive not only to the magnitude of the polarisation basis states, but also their relative phase. Polarisation compensation therefore consists of ensuring a well-behaved input-output relation for every possible input polarisation, not just a single one. This is equivalent to satisfying two simultaneous minimisation problems for two non-orthogonal polarisations, as opposed to a minimisation problem for a single polarisation state. While PM fibres, despite the name, are unsuited for preserving general polarisation states as their strong birefringence scrambles the phase between the two polarisation eigenstates of the fibre, a variety of techniques typically based on, or incorporating active feedback control, have been developed to tackle the polarisation compensation problem in SM fibres~\cite{chen2007active,xavier2008full,chen2009stable,pan2017improved,shi2021fibre,ramos2022full,peranic2022polarization}.

In laboratory-scale optical setups, however, active feedback control on the polarisation is often not employed due to cost and complexity reasons, or because the time scale of the polarisation fluctuations is long enough that initial post-alignment corrections to the polarisation suffice. Since these corrections are typically either done manually, or with low-bandwidth mechanical components, it is desirable to have a prescriptive method for compensating the polarisation. When using fibre polarisation controllers this is complicated by the fact that the input state to the controller is both random and unknown, and the degrees of freedom (DOFs) of the polarisation controller therefore map to the output polarisation state in an unpredictable way. Practically, this means that all the controller DOFs couple to both minimisation problems, making it challenging to simultaneously satisfy both of them. This problem is exacerbated by the fact that the two conditions are typically not measured in parallel. Here we review a method to decouple the two minimzation problems, thereby allowing for a simple step-by-step procedure for polarisation compensation.
\section{The method}
As the method presented here is particularly suited to quantum optics laboratories, we will use the language of quantum mechanics to describe the polarisation states, letting $\ket{\cdot}$ represent a single-photon or coherent state with a given polarisation. This description is equivalent to a formulation using Jones calculus in classical optics.

\begin{figure*}
\centering
\includegraphics[width=\linewidth]{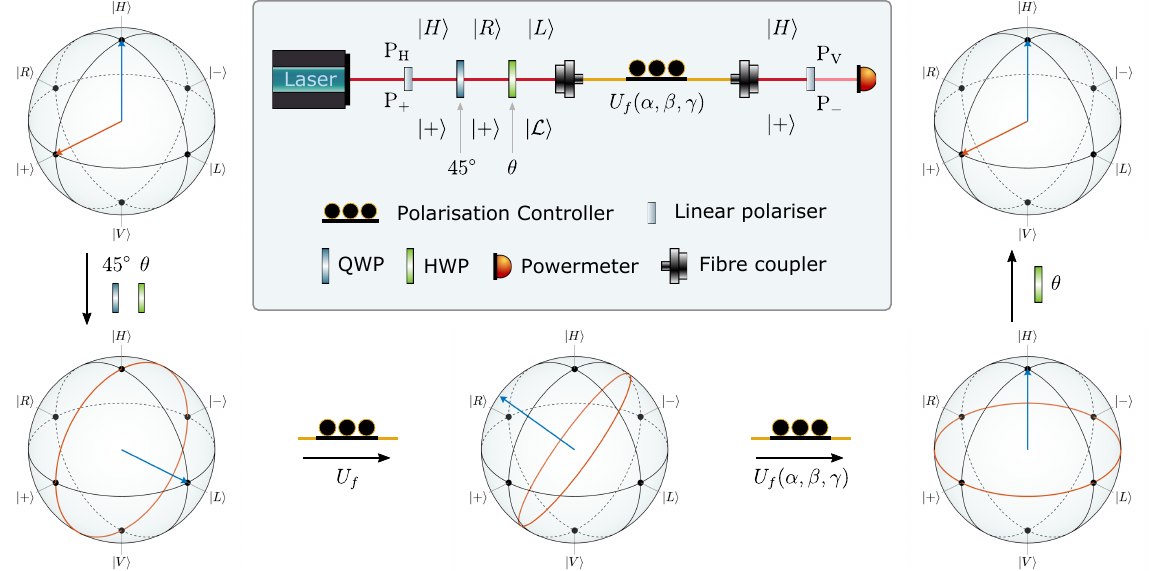}
\caption{(Top, centre) Minimal optical setup for the polarisation compensation. The labels above and below the beam path indicate the evolution of the two input polarisation states throughout the setup after the compensation has been performed. Here $\ket{\mathcal{L}}$ refers to a general linearly polarised state, not to be confused with the left-handed circularly polarised state $\ket{L}$. (Top, left) The two input states $\ket{H}$ (blue) and $\ket{+}$ (red) before the two wave plates, drawn on the Poincaré sphere. (Bottom, left) Polarisation states after the two wave plates and immediately before the fibre. This figure represents the image of the two states for all possible choices of HWP angle $\theta$. The state $\ket{H}$ has been mapped to $\ket{L}$, and $\ket{+}$ has been mapped to a state somewhere on the great circle of linearly polarised states. (Bottom, centre) Without compensation the input states get mapped to random states on the Poincaré sphere by the fibre transformation $U_f$. (Bottom, right) Using the fibre polarisation controller the input state $\ket{H}$ is mapped to itself after the fibre, and $\ket{+}$ is mapped to the equator of the Poincaré sphere. (Top, right) By tuning the angle $\theta$ of HWP before the fibre the ellipticity angle of the image of $\ket{+}$ can be set to zero, thereby ensuring that the identity transformation is realised. \label{fig:setup}} 
\end{figure*}

The polarisation compensation problem under consideration consists of ensuring that all polarisation states are mapped to themselves after propagating through a fibre, or in other words that the total polarisation transformation of the fibre should be the identity transformation. This condition can be satisfied by ensuring that two fixed non-orthogonal states are correctly transformed. For example:
\begin{equation}
    \ket{H} \xmapsto{U_f} \ket{H}, \qquad
    \ket{+} \xmapsto{U_f} \ket{+},
\end{equation}
where $U_f$ is the fibre tranformation, $\ket{H}$ is a horizontally polarised state, $\ket{+} = (\ket{H}+\ket{V})/\sqrt{2}$ is a diagonally polarised state, and $\ket{V}$ is a vertically polarised state. Experimentally, this is done by using a polariser at the fibre input to prepare the state $\ket{H}$ ($\ket{+}$) and minimising the transmission through an orthogonal polariser $\ketbra{V}{V}$ ($\ketbra{-}{-}$) after the fibre, thereby ensuring
\begin{align}
\label{eq:hmin}
\lvert
\langle V|U_f(\alpha,\beta,\gamma)|H\rangle
\rvert^2 &\approx 0
,\\
\label{eq:pmin}
\lvert
\langle -|U_f(\alpha,\beta,\gamma)|+\rangle
\rvert^2 &\approx 0,
\end{align}
where $\alpha$, $\beta$, $\gamma$ are free parameters of the polarisation controller for the fibre. If these parameters are experimentally adjusted such that \eqref{eq:hmin} is satisfied, then attempting to subsequently satisfy \eqref{eq:pmin} will generally break the condition \eqref{eq:hmin}. As illustrated in Fig.~\ref{fig:setup}, this problem can be solved by using two wave plates, placed before the fibre, to introduce an additional degree of freedom that is decoupled from the condition \eqref{eq:hmin}. Specifically, a quarter-wave plate (QWP) fixed at \SI{45}{\degree} from the vertical axis, followed by a half-wave plate (HWP) at a variable angle $\theta$. The action of these wave plates is described by the following unitary transformations:
\begin{align}
\mathrm{QWP}(\varphi) &= \mathrm{exp}\bigl[-i\frac{\pi}{4}(\sin (2\varphi) \sigma_x + \cos (2\varphi) \sigma_z)\bigr],\\
\mathrm{HWP}(\theta) &= \mathrm{exp}\bigl[-i\frac{\pi}{2}(\sin (2\theta) \sigma_x + \cos (2\theta) \sigma_z)\bigr],
\end{align}
where
\begin{equation}
    \sigma_x =
    \begin{bmatrix*}
        0 & \phantom{-}1 \\
        1 & \phantom{-}0
    \end{bmatrix*}, \qquad
    \sigma_z =
    \begin{bmatrix*}[r]
        1 & 0 \\
        0 & -1
    \end{bmatrix*}.
\end{equation}
Here the convention that $\ket{H}$ ($\ket{+}$) is the eigenvector corresponding to the positive eigenvalue of $\sigma_z$ ($\sigma_x$) is used. The quarter-wave plate at \SI{45}{\degree} transforms the horizontally-polarised state to right-handed circularly polarised light ($\ket{R} = (\ket{H}-i\ket{V})/\sqrt{2}$), and leaves the diagonally-polarised light unchanged:
\begin{equation}
\label{eq:qwpmap}
    \mathrm{QWP}(\SI{45}{\degree})\ket{H} = \ket{R},\qquad
    \mathrm{QWP}(\SI{45}{\degree})\ket{+} = \ket{+}.
\end{equation}
The half-wave plate then, independently of its angle, flips the handedness of the circularly polarised light, and rotates the diagonally polarised light to another linear polarisation:
\begin{alignat}{3}
    \label{eq:hbeforefiber}
    &\mathrm{HWP}(\theta)\ket{R} &&= \ket{L},\\
    &\mathrm{HWP}(\theta)\ket{+} &&=
    \sin(2\theta)\ket{+} + \cos(2\theta)\ket{-}.
\end{alignat}
Note that, as will become relevant shortly, the transformed diagonally polarised state can equivalently be expressed in terms of the circularly polarised states:
\begin{equation}
\label{eq:ciruclarbasis}
    \mathrm{HWP}(\theta)\ket{+} =\frac{1}{\sqrt{2}}(\ket{L}+e^{i(4\theta-\pi/2)}\ket{R}).
\end{equation}
Using \eqref{eq:qwpmap} and \eqref{eq:hbeforefiber} the first minimisation condition reads
\begin{equation}
\label{eq:hmin2}
\begin{aligned}
\lvert
\langle V|U_f(\alpha,\beta,\gamma)\mathrm{HWP}(\theta)\mathrm{QWP}(\SI{45}{\degree})|H\rangle
\rvert^2 &\approx 0\\
\iff
\lvert \langle V|U_f(\alpha,\beta,\gamma)|L\rangle
\rvert^2 &\approx 0
\end{aligned}
\end{equation}
and is independent of the HWP angle $\theta$. Satisfying this condition implies that
\begin{align}
\label{eq:hcondition}
U_f(\alpha,\beta,\gamma)\ket{L} &= e^{-i\phi/2}\ket{H}\\
\label{eq:vcondition}
U_f(\alpha,\beta,\gamma)\ket{R} &= e^{i\phi/2}\ket{V},
\end{align}
for some unknown phase $\phi$. The minimisation condition for the diagonal polarisation can be expressed as
\begin{equation}
\lvert
\langle -|U_f(\alpha,\beta,\gamma)\mathrm{HWP}(\theta)\mathrm{QWP}(\SI{45}{\degree})|+\rangle
\rvert^2 \approx 0.
\end{equation}
Using \eqref{eq:ciruclarbasis} together with \eqref{eq:hcondition} and \eqref{eq:vcondition} this can be re-written as
\begin{equation}
\begin{gathered}
\begin{aligned}
\frac{1}{4} \lvert
(\bra{H}-\bra{V})
(\ket{H}&+e^{i(4\theta-\pi/2+\phi)}\ket{V})
\rvert^2 \approx 0
\\
\iff
4\theta&+\phi - \pi/2 \approx 0.
\end{aligned}
\end{gathered}
\end{equation}
This condition can always be satisfied by tuning the HWP angle $\theta$, without affecting the first condition \eqref{eq:hmin2}.
The decoupling of the two conditions has a simple visual interpretation, illustrated in Fig.~\ref{fig:setup}: the image of $\ket{+}$ under the two wave plates is the great circle of linear polarisation on the Poincaré sphere, which gets mapped to a randomly oriented great circle by the fibre.
After using the fibre polarisation controller to satisfy the condition \eqref{eq:hmin2}, thereby mapping $\ket{H}$ to itself, this great circle is mapped to the equator, since the unitary rotation of the fibre preserves the \SI{90}{\degree} angle between the states.
Consequently the HWP angle $\theta$ is mapped to the ellipticity angle $\chi$ (modulo a random offset), or equivalently the angle in the equatorial plane, which can always be chosen such that $\ket{+}\mapsto\ket{+}$.

The steps for practically carrying out the compensation using the setup illustrated in Fig.~\ref{fig:setup} are the following:
\begin{enumerate}
    \item Inject classical light into the fibre.
    \item Place a horizontal polariser before the two wave plates to prepare the state $\ket{H}$, and place a vertical polariser in front of the fibre output.
    \item Use a powermeter to measure the transmission through the vertical polariser, and minimise the optical power with the fibre polarisation controller.
    \item Replace the polariser before (after) the fibre with a diagonal (anti-diagonal) polariser.
    \item Use the half-wave plate to minimise the transmission through the measurement polariser.
    \item Put back the horizontal (vertical) polariser to re-check the polarisation contrast, and repeat steps 3-6 if necessary.

\end{enumerate}

\begin{figure*}
\centering
\includegraphics[width=\textwidth]{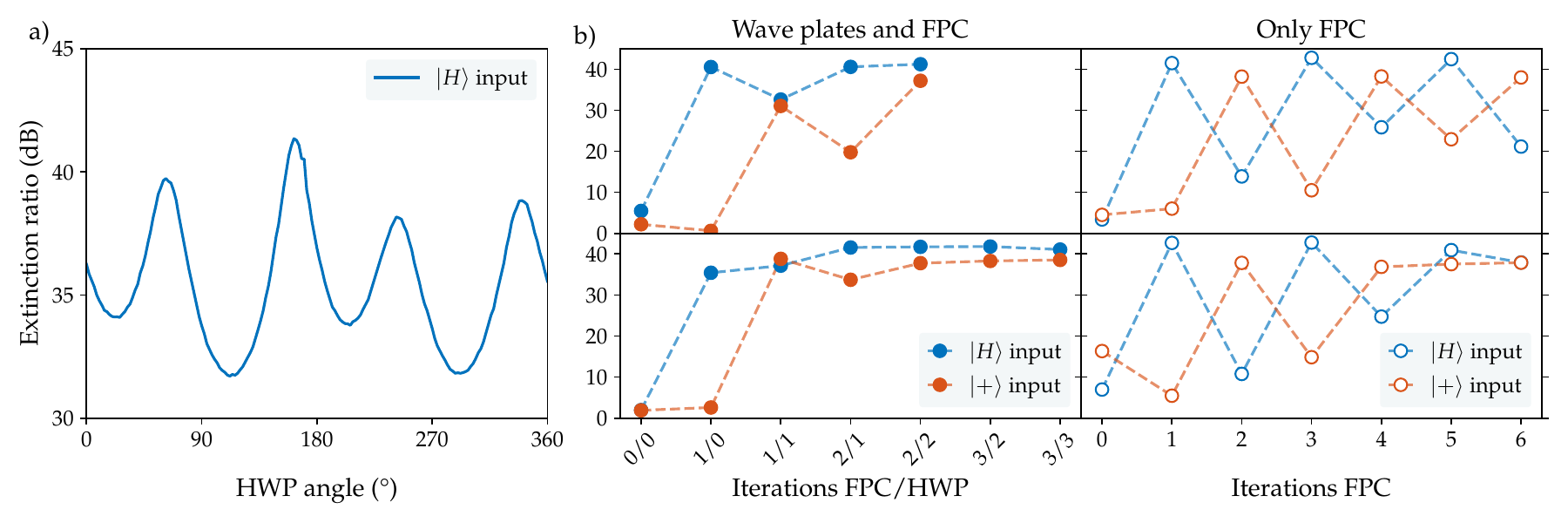}\vspace{-0.1cm}
\caption{\textbf{a) }In the presence of imperfections rotating the HWP slightly affects the state $\ket{H}$. This is illustrated above with a measurement of the polarisation contrast for the input state $\ket{H}$ after the fibre. This contrast remains well above \SI{30}{\decibel} for all HWP angles, indicating that the rejected power changes by less than one part in 1000, and the state remains almost unaffected.
\mbox{\textbf{b) } (Left, discs)}
Repeating the compensation protocol can lead to higher polarisation contrasts. This is illustrated here by showing two examples of the extinction ratio (ER) for both input polarisation states at every step of several full protocol iterations. Initially the fibre transforms the states randomly resulting in a low ER. In the first step the state $\ket{H}$ (blue) is compensated with the fibre polarisation controller (FPC). Next, the HWP is used to compensate $\ket{+}$ (red), which in the process generally slightly degrades the contrast for $\ket{H}$ due to experimental imperfections. In the third step the FPC is once again used to compensate $\ket{H}$, and after adjusting $\ket{+}$ with the HWP ERs of 41 and \SI{37}{\decibel} (42 and \SI{38}{\decibel}) are reached for $\ket{H}$ and $\ket{+}$ in the top (bottom) figures, respectively. A third repetition of the protocol does, in this case, not increase the extinction ratio further. (Right, circles) When only using an FPC, compensating one input state strongly affects the other, leading to a nondeterministic procedure that converges slowly (bottom), or not at all (top). This results in worse overall polarisation contrast. \label{fig:results}} 
\end{figure*}

We remarked earlier that due to the random polarisation state inside the fibre, the polarisation controller DOFs map to the output polarisation in an unpredictable way. This is still true in step 3 of the method described here. However, in this method the polarisation controller is only used to satisfy a single minimisation condition, and the redundant DOF of the controller means that there is not a single minimum, but rather a one-dimensional parameter space satisfying the condition. In practice, the minimisation is therefore simple to carry out.

Note that, while the description of the method here assumed the two wave plates to be on the input side of the fibre, the compensation procedure can also be carried out with light launched in the opposite direction.
\section{Practical considerations}

The decoupling of the two minimisation conditions relies on the two wave plates having the correct retardances, and the QWP being at \SI{45}{\degree} from the horizontal polariser. In practice, these conditions are never satisfied exactly, and as a consequence rotating the HWP will slightly affect the minimisation condition \eqref{eq:hmin2}. With high quality optical components and careful alignment, however, this effect can be made small. An example measurement of the polarisation contrast for the $\ket{H}$ polarisation as a function of the HWP angle $\theta$ is shown in Fig.~\ref{fig:results}(a). The contrast is defined as the extinction ratio (ER), which is the ratio between the power (P) incident on and transmitted through a polariser: $\mathrm{ER} = \mathrm{P}_{\mathrm{total}}/\mathrm{P}_{\mathrm{transmitted}}$. It can be seen that the polarisation state is largely insensitive to the rotation of the HWP. In situations where high polarisation contrast is needed, it is nevertheless prudent to re-check the ER in the $\ket{H}$/$\ket{V}$ basis after completing one round of the procedure. It has been our experience that iteratively performing two, or sometimes three rounds of compensation can result in higher simultaneous contrasts. This is illustrated in Fig.~\ref{fig:results}(b), showing the ER for both states for each step in the iterative method, as well as a comparison to only using an FPC.

To achieve high polarisation contrasts the relative angle of all polarisers is also of high importance, and it is therefore recommended to use pre-calibrated polarisers in locking rotation mounts for all four polarisation states. A vertical polariser is easily produced by extinguishing a horizontal polariser taken as reference, while a diagonal polariser can be set either by balancing the transmission through two crossed polarisers sandwiching it, or by using a high-precision motorized rotation mount to accurately rotate a polariser to \SI{-45}{\degree} from the horizontal reference, against which the diagonal polariser is then extinguished. Finally, an anti-diagonal polariser is set by extinguishing the diagonal polariser. An initial guess for the QWP angle, which should be \SI{45}{\degree}, can be found from a calibration curve of the wave plate. Better performance can sometimes be achieved by minor adjustments to the wave-plate angle. Such adjustments should be carried out by first performing steps 1-2 of the method, and then attempting to minimise the observed change in polarisation contrast when rotating the HWP. Spatial non-uniformities in the wave plate can also give rise to an asymmetry whereby rotating the QWP by \SI{180}{\degree} lowers the coupling between the two bases. Similarly, the \SI{90}{\degree} periodicity of an ideal HWP implies that \eqref{eq:pmin} should have four minima, but in practice one might result in a higher extinction ratio than the others.

\section{Discussion}
We have recalled a prescriptive step-by-step method for fibre polarisation compensation, particularly suited for table-top quantum optics experiments, in which the two measurement bases are decoupled. This is accomplished by mapping a HWP before the fibre to a source of tunable birefringence after the fibre. The compensation procedure could therefore be also carried out with a direct source of tunable birefringence, such as electro-optic devices, or strongly birefringent optics with a tilt degree of freedom. The former have drawbacks in terms of cost and complexity, while the latter can cause unwanted beam distortions due to non-normal incidence or spatial walk-off. Tunable birefringence can alternatively be achieved with birefringent wedge pairs, while maintaining near-normal incidence and small walk-off. Setting aside the fact that wave plates are a more common optical element and thus more likely to be on hand, this approach has minor practical drawbacks, requiring a more onerous pre-alignment and more complex opto-mechanical components.

Compared to only using a single fibre polarisation controller the method presented here requires two additional optical elements, but is significantly easier to carry out or automate, leading to higher achieved polarisation contrasts in practical scenarios. The higher performance is also related to the fact that mechanical fibre polarisation controllers based on spooled fibres~\cite{lefevre1980single} typically exhibit non-negligible retardance errors, causing them to not span the full space of polarisation rotations. When only using the fibre polarisation controller to satisfy one of the minimisation conditions such imperfections do not have as adverse an effect on the device performance. This is because  an ideal three-parameter polarisation controller is overcomplete for this task, and the redundant DOF can be used to compensate for the imperfections. Finally, we conclude by reiterating that, while it was briefly touched on in~\cite{prevedel2009experimental}, the exact origins of this method remain unknown to us, and we encourage the knowledgeable reader to inform us about earlier instances of it being used.

\bibliography{sample}

\end{document}